\documentclass[pdflatex, sn-nature, referee]{sn-jnl}

\usepackage{graphicx}%
\usepackage{multirow}%
\usepackage{amsmath,amssymb,amsfonts}%
\usepackage{amsthm}%
\usepackage{mathrsfs}%
\usepackage[title]{appendix}%
\usepackage{xcolor}%
\usepackage{textcomp}%
\usepackage{manyfoot}%
\usepackage{booktabs}%
\usepackage{algorithm}%
\usepackage{algorithmicx}%
\usepackage{algpseudocode}%
\usepackage{listings}%
\usepackage{upgreek}     % add greek rm letters 
\newcommand{\be}{\begin{equation}}
\newcommand{\ee}{\end{equation}}
\newcommand{\bea}{\begin{eqnarray}}
\newcommand{\eea}{\end{eqnarray}}

\def\g{\gamma}
\def\G{\Gamma}
\def\d{\delta}

\def\e{\epsilon}

\def\n{\nu}

\def\r{\rho}

\def\t{\tau}

\def\w{\omega}

\def\blk{{\mathbf k}}

\def\blq{{\mathbf q}}

\def\blD{{\mathbf D}}

\def\blU{{\mathbf U}}

\def\callP{\mbox{$\mathcal{P}$}}

\def\Re{{\rm Re}}

\def\1op{\hat{\mathbbm{1}}}

\def\AA{\mathring{\mathrm{A}}}

\begin{document}

% \title{Ultrafast Excitonic Mott transition without population inversion in WSe$_{2}$ monolayer}
%\title{Real-time dynamics of the exciton Mott transition: Dominance of screening over population inversion}
%\title{Excitonic Mott transition without population inversion}

\title{%
  \centering
  Excitonic Mott transition\\[-0.8em]
  without population inversion \vspace{1.5em}
}


\author[1,2]{\sur{Oleg Dogadov}}%\email{iauthor@gmail.com}

\author[1]{\sur{Armando Genco}}%\email{iiiauthor@gmail.com}

\author[3]{\sur{Allison R. Cadore}}%\email{iiiauthor@gmail.com}

\author[3]{\sur{James A. Kerfoot}}%\email{iiiauthor@gmail.com}

\author[3]{\sur{Evgeny M. Alexeev}}%\email{iiiauthor@gmail.com}

\author[3]{\sur{Osman Balci}}%\email{iiiauthor@gmail.com}

\author[1]{\sur{Chiara Trovatello}}%\email{iiiauthor@gmail.com}

\author[4]{\sur{Kenji Watanabe}}%\email{iiiauthor@gmail.com}

\author[4]{\sur{Takashi Taniguchi}}%\email{iiiauthor@gmail.com}

\author[5]{\sur{Seth Ariel Tongay}}%\email{iiiauthor@gmail.com}

\author[3]{\sur{Andrea C. Ferrari}}%\email{iiiauthor@gmail.com}

\author[1,6]{\sur{Giulio Cerullo}}%\email{iiauthor@gmail.com}

\author*[1]{ \sur{Stefano Dal Conte}}\email{stefano.dalconte@polimi.it}

\author*[7,8]{ \sur{Gianluca Stefanucci}}\email{stefanuc@roma2.infn.it}

\author*[7,8]{ \sur{Enrico Perfetto}}\email{eperfett@roma2.infn.it}

\affil[1]{Department of Physics, Politecnico di Milano, Piazza Leonardo da Vinci 32, 20133 Milan, Italy}

\affil[2]{Department of Physical Chemistry, Fritz Haber Institute of the Max Planck Society, 14195 Berlin, Germany}

\affil[3]{Cambridge Graphene Centre, University of Cambridge, Cambridge CB3 0FA, U.K.}

\affil[4]{Research Center for Materials Nanoarchitectonics, National Institute for Materials Science, 1-1 Namiki, Tsukuba 305-0044, Japan}

\affil[5]{School for Engineering of Matter, Transport and Energy, Arizona State University, Tempe, Arizona 85287, United States}

\affil[6]{IFN-CNR, Department of Physics, Politecnico di Milano, Piazza Leonardo da Vinci 32, 20133 Milan, Italy} 

\affil[7]{Dipartimento di Fisica, Universit{\`a} di
Roma Tor Vergata, Via della Ricerca Scientifica 1,
00133 Rome, Italy}

\affil[8]{INFN, Sezione di Roma Tor Vergata, Via della Ricerca Scientifica
1, 00133 Rome, Italy}


\abstract{
Exciton dissociation via the excitonic Mott transition (EMT) governs the high-density optical response of semiconductors and sets fundamental limits for optoelectronic devices.
The EMT is conventionally linked to the onset of population inversion and the emergence of optical gain.
Here, we demonstrate that this paradigm can break down under ultrafast non-equilibrium excitation. 
% Using femtosecond pump–probe spectroscopy, we drive monolayer WSe$_{2}$ into a dense photoexcited state in which the excitonic resonance is completely quenched within $\sim 100$ femtoseconds, yet optical gain is entirely absent across the explored fluence range. 
Using femtosecond pump–probe optical spectroscopy, we drive a monolayer transition metal dichalcogenide into a dense photoexcited state in which the excitonic resonance is completely quenched within $\sim \,$100~fs, while the optical gain is entirely absent across the explored fluence range. 
% Exciton dissociation thus precedes both carrier thermalization and population inversion.
State-of-the-art real-time \textit{ab initio} simulations reveal that the EMT is governed by an interplay of strongly nonthermal carrier populations and nonequilibrium  dynamical screening of the Coulomb interaction. 
% The excellent quantitative agreement between theory and experiment identifies a distinct pathway to exciton ionization that lies beyond quasi-equilibrium descriptions, establishing that population inversion is not a universal prerequisite but rather a limiting paradigm for EMT in two-dimensional semiconductors.
The quantitative agreement between theory and experiment identifies a distinct, ultrafast pathway to exciton ionization beyond quasi-equilibrium descriptions and demonstrates that population inversion is not a universal prerequisite for the EMT.

% Spatial confinement and reduced dielectric screening 
% make single-layer transition metal dichalcogenides (TMD) 
% an ideal platform for studying excitons and their many-body
% interactions in real time. Although exciton dynamics have been
% extensively investigated in the low-excitation regime, their 
% dynamical behavior under high photocurrent density, crucial 
% for applications in nanophotonics and optoelectronics,
% remains poorly understood. In this work, we use femtosecond 
% pulses to photoinject a high carrier density in monolayer 
% WSe$_{2}$, driving the system into the electron-hole plasma
% regime beyond the exciton Mott transition. The excitonic
% resonance is completely quenched, and no optical gain 
% indicative of population inversion is observed in the 
% fluence range explored. Real-time simulation based on 
% non-equilibrium Green’s function reproduce the measured 
% non-equilibrium optical response and highlight the critical
% role of non-equilibrium dynamical Coulomb screening effect.
% Our work represents a step forward in the deeper understanding
% of the photoinduced exciton Mott transition in low-dimensional 
% semiconductors and in the fully ab initio modeling of 
% materials under strong photoexcitation. 
}


%%%%%%%%%%%%%%%%%%%%%%%%%%%%%%%%%%%%%%%%%%%%%%%%%%%%%%%%%%%%%%%%%%%%%%%%
%%%%%%%%%%%%%%%%%%%%%%%%%%%%%%%%%%%%%%%%%%%%%%%%%%%%%%%%%%%%%%%%%%%%%%%%
%%%%%%%%%%%%%%%%%%%%%%%%%%%%%%%%%%%%%%%%%%%%%%%%%%%%%%%%%%%%%%%%%%%%%%%%
%%%%%%%%%%%%%%%%%%%%%%%%%%%%%%%%%%%%%%%%%%%%%%%%%%%%%%%%%%%%%%%%%%%%%%%%

\maketitle

\section*{Main} \label{sec1}

Excitons, Coulomb-bound states of electrons and holes, govern the optical response of semiconductors and insulators across a wide range of energies and length scales. Understanding how these composite quasiparticles form, propagate, and ultimately dissociate is a central problem in many-body physics and optoelectronics. 
This challenge becomes especially critical at high photoexcited carrier densities, where strong correlations and phase-space filling drive a fundamental reorganization of the excitonic system. 
Under such conditions, a semiconductor may undergo an excitonic Mott transition (EMT) \cite{PhysRevB.7.1508}, in which a gas of bound excitons destabilizes in favour of a metallic electron–hole plasma, in close analogy with the Mott transition in correlated electronic systems~\cite{Mott1968}.
The EMT has been investigated for decades in bulk semiconductors and quantum wells, and more recently in two-dimensional transition metal dichalcogenides (TMDs), where reduced dielectric screening and quantum confinement yield exceptionally large exciton binding energies \cite{WangRevModPhys}. 
Experimentally, the signature of EMT includes the progressive quenching of excitonic photoluminescence \cite{Kappei2005} or absorption \cite{chernikov2015}, the emergence of plasma-like optical responses \cite{Nagai2002}, and characteristic mid-infrared or THz signatures distinguishing bound excitons from free carriers \cite{Koch2006, Kaindl2003, Suzuki2012}.
From a theoretical perspective, the EMT has traditionally been understood within a quasi-thermal framework as the cooperative interplay between phase-space filling (Pauli blocking), bandgap renormalization (BGR), and static screening of the electron–hole (\textit{e-h}) attraction by thermally distributed carriers. 
Within this picture, the excitonic resonance is pushed toward the \textit{e-h} continuum, eventually merging with it and leading to exciton dissociation \cite{PhysRevB.7.1508, haug2009, asano2014excitonmott, PhysRevB.80.155201, Steinhoff2017}.
% The EMT sets in once population inversion is established -- strong Pauli blocking -- naturally giving rise to optical gain \cite{haug2009, dassarma2000manybody, meckbach_giant_2018, asano2014excitonmott, meckbach_ultrafast_2020}.
Once the population inversion, which naturally gives rise to optical gain, is established, the EMT occurs \cite{haug2009, dassarma2000manybody, meckbach_giant_2018, asano2014excitonmott, meckbach_ultrafast_2020}.
Consistent with this quasi-thermal picture, time-resolved optical studies performed on strongly photoexcited WS${}_2$ revealed pronounced exciton bleaching accompanied by negative absorption, or optical gain, below the optical gap \cite{chernikov2015}. While the optical gain has been observed in bilayer WS${}_2$, it appears to be less pronounced or completely absent in monolayer (1L) TMDs \cite{xu2025roomtempeature}. This difference likely arises from the direct bandgap nature of 1L-TMDs, which limits photocarrier accumulation in the material.

Despite these observations, the nature of the EMT remains poorly understood, especially away from quasi-thermal conditions. 
A strong ultrafast photoexcitation drives a system into a highly non-equilibrium state, in which multiple many-body processes, such as non-equilibrium screening, BGR, or electron–phonon scattering, arise on comparable timescales, \textit{i.e.}, well below hundreds of femtoseconds. 
In this regime, carrier distributions are highly nonthermal, leading to a retarded and weakened \textit{e-h} attraction.
The well-established quasi-thermal framework fails to capture the essential aspects of such nonequilibrium dynamics, raising the possibility that EMT may occur beyond the conditions under which it is currently understood. 
In this study, we study this highly non-equilibrium regime in a high quality 1L-TMD and demonstrate that EMT can occur in absence of population inversion.

We investigate the transient optical response of 1L-WSe$_2$ encapsulated in hexagonal boron nitride (hBN) following intense above-gap excitation, using broadband femtosecond transient reflectivity spectroscopy. 
The experiments are directly compared with real-time \textit{ab-initio} simulations based on the non-equilibrium Green’s function (NEGF) formalism \cite{svl_book}, which explicitly incorporates electron–phonon coupling and retardation  effects in the screened Coulomb interaction (dynamical screening) \cite{pavlyukh_time-linear_2022, perfetto_real_2022, perfetto_real_time_2023}.
% Immediately after photoexcitation, the A-exciton resonance undergoes a rapid loss of spectral weight and disappears entirely within $\sim \,$100~fs, signaling complete exciton ionization. 
% % Strikingly, this ultrafast ionization occurs in the absence of optical gain -- carrier occupations near the band edges remain far from population inversion -- and precede carrier thermalization. 
% Strikingly, this ultrafast ionization occurs in the absence of optical gain and precedes carrier thermalization. 
% For all time delays, the occupations near the band edges remain far from population inversion.
% The resulting exciton-free, gain-free phase persists for several picoseconds, before excitonic features gradually recover.
% The quantitative agreement between experiment and theory establishes that the transient EMT in 1L-WSe$_{2}$ is governed by the interplay of dynamical screening and strongly nonthermal carrier populations. 
% We further show that orthodox quasi-thermal treatments invariably predict pronounced optical gain during the transition, in stark contrast to our observation. 
% Our results demonstrate that population inversion is not a universal prerequisite for exciton dissociation, but rather a limiting paradigm applicable only under quasi-equilibrium conditions.
A complete disappearance of the A exciton resonance for pump fluences above the EMT threshold, together with strong BGR and the absence of optical gain, observed in the experiment and predicted by our simulations, establish that the transient EMT in 1L-WSe$_{2}$ is governed by the interplay of dynamical screening and strongly nonthermal carrier populations. 
We further show that standard quasi-thermal treatments invariably predict pronounced optical gain during the transition, in stark contrast to our observations. 
Our results demonstrate that population inversion is not a universal prerequisite for exciton dissociation, but rather a limiting paradigm applicable only under quasi-equilibrium conditions.

\begin{figure}[h]
    \centering
    \includegraphics[width=1.0\textwidth]{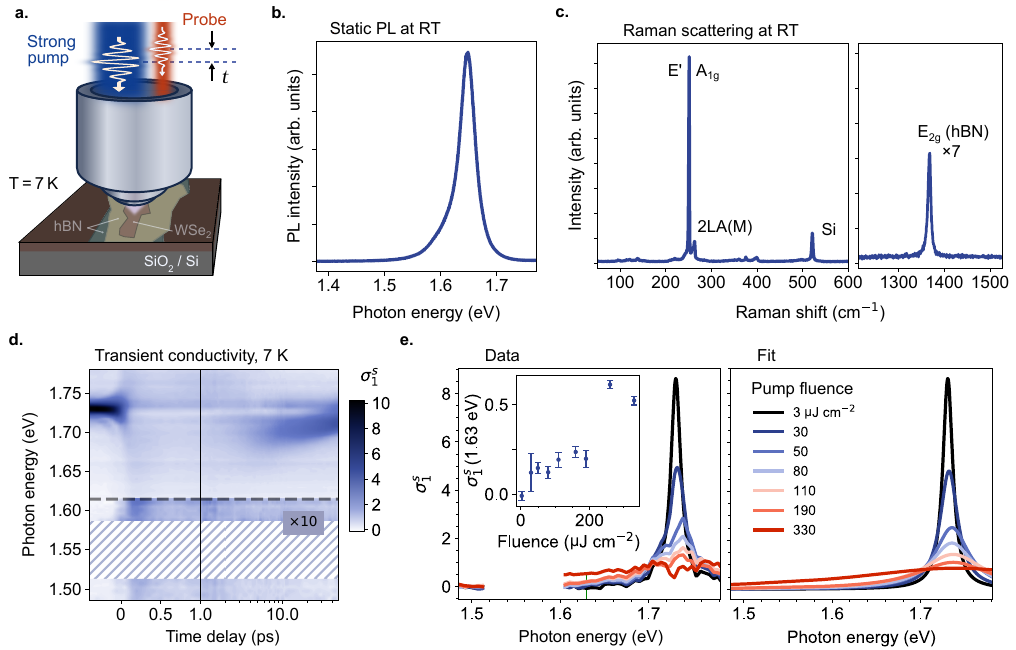}
    \caption{
    \textbf{Pump-probe measurements of the EMT in 1L-$\mathbf{WSe_2}$.} 
    (\textbf{a}) Sketch of the hBN-encapsulated 1L-$\mathrm{WSe_2}$ measured
	in the pump-probe microscope (not to scale).
    (\textbf{b}--\textbf{c}) Sample characterization at room temperature (RT):
    (\textbf{b}) PL spectrum and (\textbf{c}) Raman spectrum. 
    (\textbf{d}) Pseudo-color energy-time conductivity map for high pump fluence
	(330~$\upmu \mathrm{J \, cm^{-2}}$). 
    The map displays the disappearance of the excitonic peak shortly after the photoexcitation. 
    (\textbf{e}) Fluence dependence. 
    Left: Transient conductivity spectra at 400~fs time delay for selected pump fluences. 
    In the energy region around 1.55~eV, the probe is eliminated by a filter.
    The vertical line at 1.63~eV indicates the energy for which the $\sigma_1^s$ fluence dependence is shown in the inset.
    Right: corresponding fit with Lorentzian profile.
    Sheet conductivity is reported in $\frac{2e^2}{h}$ units.
    }
    \label{fig:exp}
\end{figure}

The experiments in this work are performed on a high-quality hBN-encapsulated 1L-$\mathrm{WSe_2}$, fabricated by the exfoliation method. 
The sample is placed on a highly reflective silicon substrate terminated with a 90-nm $\mathrm{Si O_2}$ layer, as illustrated in Figure~\ref{fig:exp}a (see Methods and Supplementary Note~1 for details). 
Steady-state photoluminescence (PL) and Raman spectra of the sample acquired at room temperature are reported in Figure~\ref{fig:exp}b--c. 
The PL spectrum shows a single peak corresponding to the A exciton 1$s$ state.
The frequencies of the observed Raman modes correspond well to the expected values for 1L-WSe${}_2$ and multilayer hBN \cite{reich2005resonant, barbone2018charge}.
We further characterize low-temperature optical properties of the sample by reflectance contrast (RC) measurements. 
We define RC spectrum as $\mathrm{RC} = (R_\mathrm{sub} - R_0) / R_\mathrm{sub}$, where $R_0$ is reflectance spectrum of the sample and $R_\mathrm{sub}$ is measured on a spot with only the hBN flakes on the substrate. 
In the RC measurements (see Supplementary Figure~3) a single narrow and intense peak associated with the absorption of the $1s$ state of A exciton is observed at 1.73~eV.

The transient reflectivity measurements are performed at 7~K in a home-built pump-probe confocal microscope, equipped with a closed cycle helium cryostat \cite{genco2022}, as illustrated in Figure~\ref{fig:exp}a (see details in Methods and Supplementary Note~2).
The sample is excited above the bandgap with narrow-band linearly polarized (10~nm) $\sim \, $150~fs pulses at 2.53~eV, generated in a noncollinear optical parametric amplifier (NOPA). 
The transient dynamics is monitored with a supercontinuum (SC) probe covering the A exciton peak and the lower energy region, where both the BGR and the possible onset of optical gain are expected. 
The pump and the probe beams are focused to ca.~20~$\upmu$m and 5~$\upmu$m spots, respectively, to ensure homogeneous excitation of the probed area. 
The measured static RC and differential reflectivity $\Delta R (E, \, t) / R_0$, where $E$ is the probe photon energy and $t$ is the pump-probe delay, is used to extract the temporal evolution of the sheet optical conductivity $\sigma^s (E, \, t)$ by performing Kramers-Kronig constrained analysis, as explained in the Methods (see also Supplementary Note~3). 

Figure~\ref{fig:exp}d reports a pseudo-color map of the nonequilibrium conductivity as a function of probe photon energy and pump-probe delay.  
The energy range between 1.51~eV and 1.59~eV is blocked by a notch filter, since in this region the probe spectrum is significantly perturbed by the laser fundamental used to generate the SC. 
%The measurement is performed 
%at 7~K 
The data are acquired at a pump fluence of 330~$\mathrm{\upmu J \, cm^{-2}}$.
Here and throughout the text, the real part of sheet conductivity is reported in $\frac{2e^2}{h}$ units.
Before time zero, \textit{i.e.}, before the sample is perturbed by the pump, the spectrum is dominated by a single positive peak at around 1.73~eV, corresponding to the A exciton 1$s$ transition. 
Upon intense photoexcitation, the exciton peak initially displays a red-shift and broadens, and eventually disappears completely within ca.~100~fs, indicating that the system undergoes an EMT.
At the same time, an increase of optical conductivity is observed in the whole probe range above and below the excitonic resonance, with no signature of optical gain, which would result in a negative optical conductivity.
The low energy part of the map in Figure~\ref{fig:exp}d is multiplied by a factor of ten for clarity. On a picosecond timescale (ca.~10~ps), the excitonic transient signal recovers as the photoexcited carrier density falls below the Mott threshold through recombination and diffusion, allowing excitons to reform.
%The transient signal persists for several picoseconds, until the exciton peak starts to reappear at lower energies after ca.~10~ps. 

The fluence dependence of the optical conductivity recorded at $\sim \, $400~fs time delay along with the corresponding fit with a single Lorentzian profile is presented in Figure~\ref{fig:exp}e.
As the fluence increases, the exciton peak becomes broader and its magnitude decreases, until it gets almost completely quenched above ca.~150~$\mathrm{\upmu J \, cm^{-2}}$. 
For higher pump fluences, a broad featureless band is observed, and the signal below 1.70~eV is markedly increased, indicating an increment of low energy absorption due to a significant BGR. 
This behavior is clearly illustrated by the inset in Figure~\ref{fig:exp}e, showing the evolution of $\sigma_1^s$ 100~meV below the A exciton 1$s$ peak. 
% This behavior is clearly illustrated by the peak width fluence dependence, shown in the inset of Figure~\ref{fig:exp}e: 
% As the pump fluence increases, the peak width gets progressively larger until it reaches a plateau at approximately 150~$\mathrm{\upmu J \, cm^{-2}}$, where a complete quenching is observed. 
% At higher fluences, a very broad absorption feature appears. 
Above $\sim \, $250~$\mathrm{\upmu J \, cm^{-2}}$ the transient signal saturates and remains unchanged upon further increasing the pump fluence, until sample damage is observed. 

The photoinduced changes in the optical properties of the system indicate that the EMT in the studied material is due to high photocarrier density, yet its microscopic mechanism is not related to population inversion. 
To shed light onto the mechanism, we perform real-time simulations based on the \textit{ab-initio} NEGF method of Ref.~\cite{perfetto_real_time_2023} (see also Methods and Supplementary Note~4 for details).
Nonequilibrium dynamical screening is treated at the level of the $GW$ self-energy, where the retarded screened interaction $W(t,t')$  plays a central role, as the EMT develops during the build-up of the screening, as discussed below. 
To accurately describe carrier and phonon relaxation, electron-phonon scattering is included through the Fan-Migdal (FM) self-energy. 
Importantly, the NEGF framework ensures a consistent exchange of energy between electrons and phonons, guaranteeing total energy conservation throughout the simulations.    
Our approach unifies, within a single framework, population dynamics, typically described by Boltzmann or semiconductor Bloch equations, with the Bethe–Salpeter equation formalism, used to compute the nonequilibrium spectra. 
It further surpasses these methods by avoiding the Markov approximation describing memoryless systems\cite{stefanucci_semiconductor_2024}, which breaks down during the early transient, and by incorporating fully dynamical rather than strictly static screening \cite{haug2009, Hannewald2004, KIRA2006155, perfetto_nonequilibrium_2015, PhysRevB.94.245303, PereaCausin2020, Lohof2019}.   

\begin{figure}[h]
    \centering
    \includegraphics[width=1.0\textwidth]{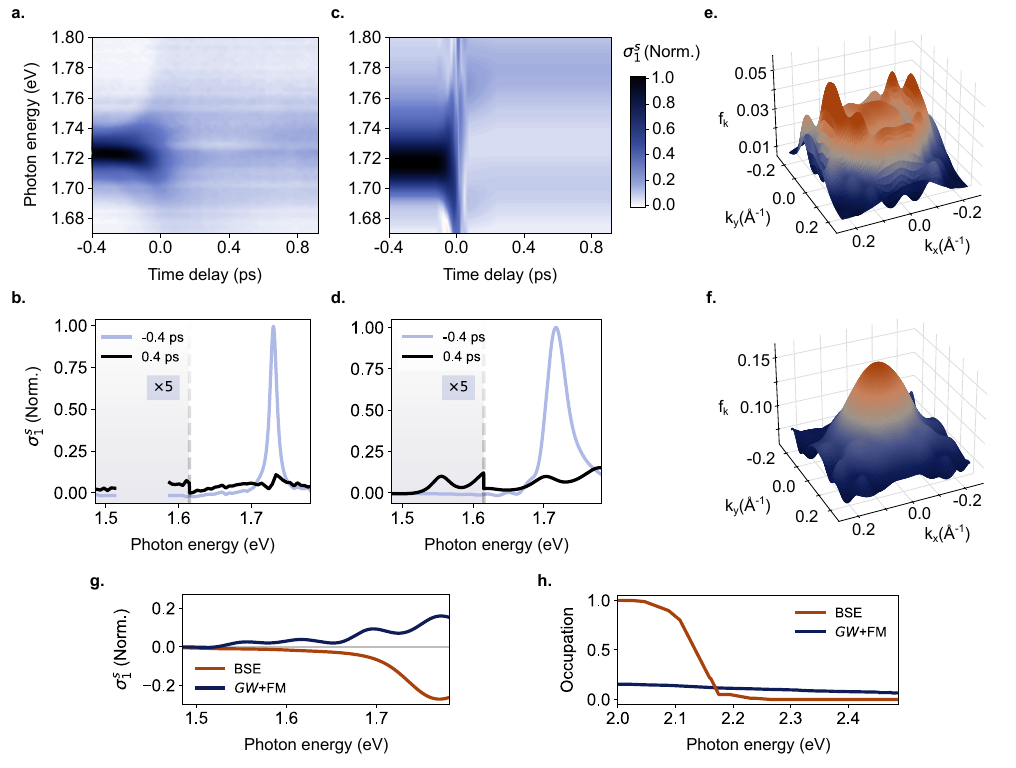}
    \caption{
    \textbf{Sub-picosecond dynamics.}
    Experimental (\textbf{a}) and simulated (\textbf{c}) transient
	conductivity maps. 
    Transient spectra at select time delays corresponding to experimental
	(\textbf{b}) and simulated (\textbf{d}) maps. 
    The spectra show a disappearance of the excitonic peak upon photoexcitation
	and an increase of absorption at lower energies.
    Simulated spin-averaged density profiles right after photoexcitation (0~fs) (\textbf{e})
	and at the  
	the Mott transition ($30$~fs) (\textbf{f}),
	showing migration of photocarriers towards the conduction band 
	minimum. Momenta are measured relative to the $K$ point,
	which defines the  origin.
    Comparison between spectra (\textbf{g}) and spin-averaged populations (\textbf{h})
	in different approaches, namely the NEGF method with self-energy in
	the \mbox{$GW$+FM} and the BSE with thermal
	carriers at temperature 100~K. In \textbf{h} the energy of 
	carrier occupations are referred to the top of the valence band. 
    }
    \label{fig:exp_theory}
\end{figure}

The simulations are performed by driving the 1L-WSe$_{2}$ out of equilibrium with a 100~fs pump pulse centered at $2.53~$eV with 300~$\upmu \mathrm{J \, cm^{-2}}$ fluence, corresponding to a photoexcited carrier density of approximately $1.6 \times 10^{13} \ \mathrm{cm^{-2}}$ in each of the $K/K^\prime$ valleys. %We emphasize that our simulations are restricted to electronic bands at the $K/K^\prime$ valleys and do not include fast scattering processes to energetically nearby valleys. These processes can promote carrier accumulation of photoinduced electron–hole pairs \cite{Lohof2019} and may favor optical gain, as reported in bilayer TMDs \cite{chernikov2015, xu2025roomtempeature}.  
% An excellent agreement between the measured and simulated transient conductivity is obtained, as shown in Figures~\ref{fig:exp_theory}a-d.
Figure~\ref{fig:exp_theory}a-d compares the experimental and simulated transient conductivity maps, displaying an excellent agreement. 
The numerical results provide important insights into the microscopic ultrafast dynamics.  
At maximum pump–probe temporal overlap (near zero delay), electrons and holes are distributed by the pump pulse away from the band edges, as shown in Figure~\ref{fig:exp_theory}e, while the build-up of the screening is not yet fully established (see Supplementary Figure~9). 
The transient red-shift of the A-exciton can therefore be primarily attributed to the BGR.
At longer time delays, carriers  rapidly relax toward the band edges. 
The exciton peak is progressively bleached, vanishing after several tens of femtoseconds, while the carrier population remains highly non-thermal (see Figure~\ref{fig:exp_theory}f) and the screening has not yet reached its steady-state value.

Both theory and experiment show here an increase of absorption  down to $\sim 1.5$~eV, with no evidence of optical gain. 
As the estimated direct bandgap of 1L-$\mathrm{WSe_2}$ is around  $2.0$~eV~\cite{madeo2020}, we infer a renormalization of $\sim \, $0.5~eV, consistent  with previous reports for 1L-$\mathrm{WS_2}$ \cite{chernikov2015}. 

Nonequilibrium dynamical screening plays a dual and inseparable role \cite{perfetto_real_2022, perfetto_real_time_2023}, with each component essential for reproducing the experimental observations.   
On the one hand, it weakens the electron–hole attraction; on the other hand, retardation effects open additional electron–electron scattering channels that, under strong photoexcitation, dominate over phonon-mediated scattering processes (see also Supplementary Note~10) and govern the ultrafast intravalley carrier dynamics. 
This rapid relaxation reshapes the initially photoexcited carrier population into a strongly non-thermal yet non-inverted distribution, thereby suppressing the population inversion and the emergence of the optical gain.

Figure~\ref{fig:exp_theory}g compares the conventional quasi-thermal description of the EMT, represented by the BSE spectrum at $T=100$~K, with the fully nonequilibrium $GW$+FM spectrum at a 30~fs pump–probe delay, where the Mott transition occurs.
The reference temperature of $T=100$~K approximates the heating expected after thermalization for an excitation density of $\sim \, $10${}^{13}~\mathrm{cm}^{-2}$ \cite{perfetto_real_time_2023}.
Although the total excitation density is the same in both cases, the carrier populations are distributed very differently in energy, reflecting fundamentally distinct nonequilibrium states (see Figure~\ref{fig:exp_theory}h).
In the \mbox{$GW$+FM} calculation in particular, the peak occupation (per spin) is only $\sim \, $0.13, see also Figure~\ref{fig:exp_theory}f, well below the threshold for population inversion. 
While the BSE correctly models the disappearance of the excitonic peak, consistent with the EMT, it simultaneously predicts a pronounced optical gain below the equilibrium gap.
This qualitative difference persists over a broad temperature range (0--1000~K) demonstrating the limitations of a quasi-thermal description.

\begin{figure}[tbp]
    \centering
    \includegraphics[width=1.\textwidth]{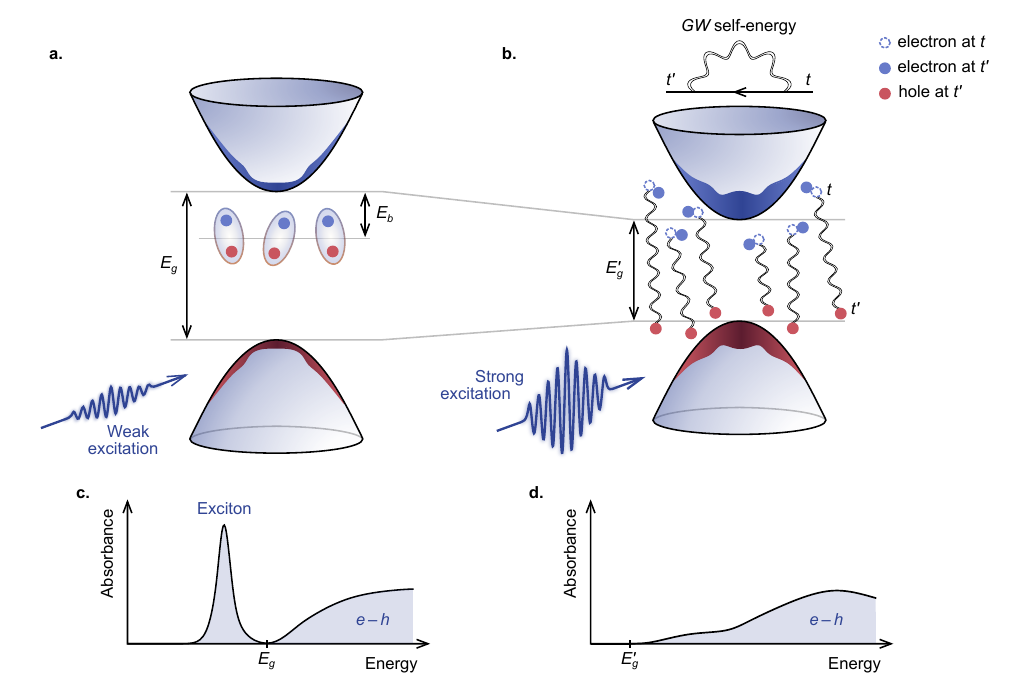}
    \caption{
    \textbf{Plasma screening of electron-hole attraction.} 
    Low (\textbf{a}) and high (\textbf{b}) excitation regimes. 
    In (\textbf{a}) the bandgap $E_{g}$ is weakly renormalized and the Coulomb attraction is effectively instantaneous. 
    Electron-hole pairs form bound excitons, with binding energy $E_{b}$. In (\textbf{b}) the exciton formation is suppressed by the reduced bandgap $E'_{g}$, nonthermal carrier populations, and nonequilibrium dynamical screening. 
    The retarded screened interaction $\d n(t',t)W$ entering the $GW$ self-energy is depicted as a double-wiggly line. 
    Schematic nonequilibrium absorption spectra in the low- (\textbf{c}) and high-density (\textbf{d}) excitation regimes are shown. 
    In (\textbf{c}) sharp excitonic resonances dominate the spectrum below $E_{g}$, marking the onset of the \textit{e-h} continuum.
	In (\textbf{d}) excitons are ionized and the \textit{e-h} continuum onset is shifted to the renormalized gap $E'_{g}$.  
	}
    \label{fig:mott}
\end{figure}

The underlying microscopic mechanism for the EMT is schematically illustrated in Figure~\ref{fig:mott}.
Following excitation of a semiconductor by a femtosecond pulse above the quasi-particle gap, photogenerated free \textit{e-h} pairs scatter towards lower-energy states. At low fluences (Figure~\ref{fig:mott}a), BGR remains weak and the effective \textit{e-h} interaction is well described by a statically screened Coulomb attraction. 
In this regime, the screened Coulomb potential supports well-defined excitonic eigenstates and the optical response is dominated by stable bound excitons \cite{Pogna2016,Sie2017,cunningham2019,Trovatello2022,calati2023} (Figure~\ref{fig:mott}c; for clarity, a single exciton 1$s$ peak is shown).
At higher pump fluences (Figure~\ref{fig:mott}b), the situation is qualitatively different. Strong BGR shifts the quasiparticle energies, while the carrier distributions remain highly non-thermal on the ultrafast timescale. 
The dense photo-generated plasma induces a time-dependent polarization that screens the Coulomb \textit{e-h} attraction dynamically rather than instantaneously.
Under these transient conditions, a hot electron at time $t'$ induces a density change $\d n(t,t')$ for all times $t>t'$, which in the static approximation reduces to $\d n(t,t')=-e^{\ast}\d(t-t')$, where $e^{\ast}$ is the statically screened electron charge. 
Thus, the interaction is governed by the dynamically screened potential $W(t,t')=\d n(t,t')v$, with $v$ the bare Coulomb  interaction~\cite{svl_book, perfetto_real_2022} and $\d n(t,t')$ the retarded density response function. 
The combined action of dynamical screening and nonthermal carrier populations drives exciton ionization without establishing population inversion (Figure~\ref{fig:mott}d).

In conclusion, we demonstrate that the EMT in semiconductors can proceed through a fundamentally nonequilibrium pathway that does not require population inversion. 
By combining femtosecond pump–probe spectroscopy with real-time \textit{ab initio} simulations, we demonstrate that intense above-gap excitation of 1L-WSe$_2$ induces complete exciton ionization within $\sim \, $100~fs, in the absence of optical gain and well before carrier thermalization. 
This behavior is strikingly different from the conventional EMT paradigm, in which exciton dissociation is expected to coincide with maximal Pauli blocking and the onset of the optical gain. 
%As shown in previous studies \cite{chernikov2015, xu2025roomtempeature}, the optical gain can be observed in bilayers of TMDs, where the indirect gap enables accumulation of photoexcited \textit{e-h} pairs.

Our results establish that, under ultrafast high-density excitation, the EMT is governed by the interplay of strongly nonthermal carrier populations and dynamical screening. 
Screening develops on the same timescale as carrier injection and relaxation, destabilizing excitonic correlations prior to the formation of a quasi-thermal electron–hole plasma. 
More broadly, these findings highlight the importance of understanding the quantum many-body dynamics of semiconductors driven far from the equilibrium, where non-thermal pathways to exciton dissociation reshape the optical response.
The resulting transformation of excitonic complexes into other quasiparticles directly impacts the absorption and emission processes that govern photonic functionalities in lasers, solar cells, and light-emitting diodes.
By clarifying the microscopic limits of operation under strong photoexcitation, this work identifies new opportunities for the light-driven control of collective excitations in quantum materials.

%%%%%%%%%%%%%%%%%%%%%%%%%%%%%%%%%%%%%%%%%%%%%%%%%%%%
%%%%%%%%%%%%%%%%%%%%%%%%%%%%%%%%%%%%%%%%%%%%%%%%%%%%

\section*{Methods} \label{sec11}

\bmhead{Sample preparation}
High-quality hBN-encapsulated 1L-$\mathrm{WSe_2}$ is fabricated at the Cambridge Graphene Centre. 
Bulk hBN crystals are grown in a barium boron nitride solvent at high pressure and temperature, as explained in \cite{casiraghi2007}.
Prepared by micromechanical cleavage of bulk crystals hBN flakes are transferred on a silicon substrate with a 90-nm $\mathrm{SiO_2}$ coverage. 
Single layers of $\mathrm{WSe_2}$ are obtained by cleavage of flux-zone grown bulk $\mathrm{WSe_2}$ crystals onto spin-coated poly-methylmetacrylate (PMMA) on glass at $80\ ^\circ$C \cite{zhang2015}. 
Encapsulated samples are fabricated using polycarbonate (PC) stamps \cite{purdie2018} after identifying suitable hBN flakes and $\mathrm{WSe_2}$ layers via optical microscopy methods \cite{taniguchi2007}. 
After PC on polydimethylsiloxane is pressed onto the top hBN flake at $40\ ^\circ$C, a 1L-$\mathrm{WSe_2}$ flake on PMMA is aligned to the hBN flake on the PC stamp and picked up at $80\ ^\circ$C. 
The bottom hBN flake is picked up at $40\ ^\circ$C. 
The encapsulated 1L-$\mathrm{WSe_2}$ is then held at $180\ ^\circ$C. 
The PC residue is removed by immersing the sample in chloroform and then ethanol for 30~minutes.

%%%%%%%%%%%%%%%%%%%%%%%%%%%%%%%%%%%%%%%%%%%%%%%%%%%%%%%%%%%%%%%%%%%%%%%%%%%%%%%%%
%%%%%%%%%%%%%%%%%%%%%%%%%%%%%%%%%%%%%%%%%%%%%%%%%%%%%%%%%%%%%%%%%%%%%%%%%%%%%%%%%

\bmhead{Experimental set-up} 
Transient reflectivity measurements are performed in a confocal microscope equipped with a closed-cycle cryostat (see Supplementary Figure~2).
A regeneratively amplified Ti-sapphire laser (Coherent, Libra) emitting 100-fs pulses at 1.55~eV at a 2~kHz repetition rate is used as a light source in the setup.
Pump pulses are generated in a home-built NOPA pumped by the second harmonic of the laser.
The NOPA signal is tuned to 2.53~eV (well above the bandgap). 
The pump is modulated by a mechanical chopper at 0.5~kHz, blocking and letting pass pairs of pulses.
The SC probe is generated by focusing the laser fundamental in a 1~mm-thick $\mathrm{Al_2 O_3}$ plate. 
The probe energy range is further selected by a set of short- and long-pass filters and attenuated by a neutral-density filter. 
The pump and probe beams are then collinearly combined by a thin beam splitter and focused on a sample inside the cryostat with an achromatic 8~mm-focal length objective lens with $\mathrm{NA} = 0.3$, mounted on a three-axis translation stage. 
The measurements are performed at 7~K. 
The reflected probe beam is then detected by a spectrometer with a high sensitivity silicon CCD.
Differential reflectivity maps are acquired by scanning the temporal delay between pump and probe pulses with a mechanical delay stage. 
The experimental set-up is further equipped with an imaging system consisting of a white light source (LED) and a CMOS camera, used to precisely align the pump and probe beams on the sample. 

%%%%%%%%%%%%%%%%%%%%%%%%%%%%%%%%%%%%%%%%%%%%%%%%%%%%%%%%%%%%%%%%%%%%%%%%%%%%%%%%%
%%%%%%%%%%%%%%%%%%%%%%%%%%%%%%%%%%%%%%%%%%%%%%%%%%%%%%%%%%%%%%%%%%%%%%%%%%%%%%%%%

\bmhead{Transient conductivity}
The optical properties of the 1L-$\mathrm{WSe_2}$ are extracted form a measured reflectivity signal by applying the Transfer Matrix Method (TMM) \cite{hecht, ivchenko}. 
In this method, the total field in the multilayer system of hBN-encapsulated 1L-$\mathrm{WSe_2}$ is calculated considering the field propagation in individual layers and applying Maxwell’s boundary conditions to establish the link between the field components in adjacent layers.
The relation between the electric and magnetic field components on the first and the last boundaries is expressed as a product of characteristic matrices of individual layers.
The transfer matrix of the system studied in this work is expressed as $M = M_\mathrm{hBN'} \, M_\mathrm{WSe_2} \, M_\mathrm{hBN''} \, M_\mathrm{SiO_2}$.
The layer parameters used in the TMM calculations are reported in Supplementary Note~3.
The dielectric function (DF) of the 1L-$\mathrm{WSe_2}$ is modeled with Kramers-Kronig constrained functions by fitting a measured reflectance contrast spectrum \cite{kuzmenko2005, ashoka2022}.
Similarly to \cite{kuzmenko2005}, the DF is first parameterized with a single Lorentzian function to correctly approximate the energy dependence outside the probed spectral range. 
Next, the modeled DF is ``corrected'' by a large number of evenly distributed model independent oscillators of fixed width in order to take into account the fine spectral details (approximately one oscillator per five experimentally measured photon energy points).
The static reflectivity of the sample $R_0(E)$ is calculated from the modeled ground state DF by employing the TMM.
The time-dependent reflectivity is then calculated from measured differential reflectivity as $R(E, \, t) = R_0 \cdot [\Delta R(E, \, t) / R_0 + 1]$.
The time-dependent DF $\hat{\epsilon}(E, \, t)$ is calculated by repeating the explained procedure for all time delays.
The transient response of the 1L-$\mathrm{WSe_2}$ is expressed in terms of the sheet optical conductivity $\sigma^s_1 (E, \, t) = \Re [\hat{\sigma}(E, \, t) \, d]$, where $\hat{\sigma} (E, \, t) = \dfrac{i \epsilon_0 E}{\hbar} [1 - \hat{\epsilon}(E, \, t)]$ is complex conductivity and $d$ is the 1L-$\mathrm{WSe_2}$ thickness.

%%%%%%%%%%%%%%%%%%%%%%%%%%%%%%%%%%%%%%%%%%%%%%%%%%%%%%%%%%%%%%%%%%%%%%%%%%%%%%%%%
%%%%%%%%%%%%%%%%%%%%%%%%%%%%%%%%%%%%%%%%%%%%%%%%%%%%%%%%%%%%%%%%%%%%%%%%%%%%%%%%%

\bmhead{NEGF framework}
The photoinduced dynamics of electrons and phonons is simulated in real-time using the NEGF method of Refs.~\cite{perfetto_real_2022, perfetto_real_time_2023}.
This method is based on the simultaneous propagation of the nonequilibrium electronic and phononic density matrices ($\rho$ and $\gamma$ respectively), together with the nuclear displacements $u_{\n}$ along the normal modes $\n$ of the material.
A key feature of the NEGF method is its linear scaling with the maximum propagation time~\cite{joost_2020, schlunzen_achieving_2020, pavlyukh_photoinduced_2021, pavlyukh2022}, while capturing non-Markovian effects arising from the double-time dependence of the non-equilibrium self-energy.
The equations of motion  can be derived from the Kadanoff-Baym equations \cite{svl_book, stefanucci2023in} using the Generalized Kadanoff-Baym ansatz~\cite{lipavsky1986, karlsson2021, pavlyukh2022}, and read
\begin{eqnarray}
    i \dot 
    \rho_{\bf{k}}(t)&=&[h^{\rm{el}}_{\bf{k}}(t),\r_{\bf{k}}(t)]-I^{\rm{el}}_{\bf{k}}(t) \nonumber \\
    i \dot \g_{\bf{k}}(t)&=&[h^{\rm{ph}}_{\bf{k}}(t),\g_{\bf{k}}(t)]-I^{\rm{ph}}_{\bf{k}}(t) \nonumber \\
    \ddot {u}_{\n} &=&-\w_{\n}^{2} u_{\n}-\sum_{\bf{k}} {\rm Tr} \{ g_{\n}(\blk,0)[\r_{\bf{k}}(t)-\r_{\bf{k}}(0)] \}\, ,  
    \label{eom}
\end{eqnarray}
where $\bf{k}$ is the crystal momentum and $\w_{\n}$ are the phonon frequencies. 
The electron density matrix $\rho_{\bf{k}}$, quasi-particle electronic Hamiltonian $h^{\rm{el}}_{\bf{k}}$, electronic collision integral $I^{\rm{el}}_{\bf{k}}$ and electron-phonon couplings $g_{\n}(\blk , \blq)$ are all matrices with electronic-band indices.
Similarly, the phonon density matrix $\g_{\bf{k}}$, free-phonon Hamiltonian $h^{\rm{ph}}_{\bf{k}}$  and phononic collision integral $I^{\rm{ph}}_{\bf{k}}$  are all matrices with normal-mode indices. The collision integrals are computed using the full density matrices $\r$ and $\g$; hence they depend on both the diagonal (populations) and off-diagonal (polarizations) elements of the matrices.    
The resulting \mbox{$GW$+FM} self-energies are evaluated using the full electronic and phononic density matrices, making them functionals of both the populations and the polarizations. 
The electronic self-energy comprises five distinct contributions, \textit{i.e.}, Hartree, statically screened exchange (SEX), Ehrenfest, $GW$ and FM, whereas the phononic self-energy in $I^{\rm{ph}}_{\bf{k}}$ constitutes of the polarization bubble diagram (see Supplementary Note~4).
The three coupled equations of motion in Eq.~(\ref{eom}) are numerically integrated by using a 4th order Runge-Kutta solver with a time step $\Delta t = 0.1$~fs as implemented in the CHEERS code~\cite{PS-cheers}.

From  the knowledge of $\r_{\bf{k}}$, $\g_{\bf{k}}$ and $u_{\n}$ we can evaluate all the observables needed to characterize the photoexcited carrier and nuclear dynamics. 
In particular, we have access to the instantaneous carrier occupations $f_{{\bf k}i}$ with momentum ${\bf k}$ in spin-band $i$, laser-induced polarizations ${\bf d}_{\bf{k}}$, phonon populations $n_{\bf{k}\n}$ with momentum ${\bf k}$ in normal-mode $\n$, and effective temperature $T$ of the lattice.
The excitonic properties are monitored by calculating the transient absorption spectrum according to~\cite{PSMS.2015} 
\be
\mathfrak{S}(\w,\t)=-2\w \, \rm{Im}[\bf{e}(\w,\t) \cdot 
\bf{d}(\w,\t)],
\ee
where $\bf{e}(\w,\t)$  is the Fourier transform of the probe field impinging the system at time $\t$ and $\bf{d}(\w,\t)$ is the Fourier transform of the total probe-induced polarization ${\bf d}(t,\t)$.
The latter is defined as ${\bf d}(t,\t)=\sum_{\bf{k}}\left[{\bf d}^{P+p}_{\bf{k}}(t,\t)-{\bf d}^{P}_{\bf{k}}(t)\right]$, where ${\bf d}^{P}_{\bf{k}}(t)$ is evaluated in the presence of the pump field only, whereas ${\bf d}^{P+p}_{\bf{k}}(t,\t)$ is evaluated in the presence of both pump  and probe fields.
The momentum-resolved dipole in the presence of an external field is defined as ${\bf d}_{\bf{k}}(t)={\rm Tr}[\blD_{\blk}\r_{\blk}(t)]$, where $\r_{\blk}(t)$ obeys Eq.~(\ref{eom}) $\blD_{\blk}$ is the dipole matrix element (see next Method Section).

%%%%%%%%%%%%%%%%%%%%%%%%%%%%%%%%%%%%%%%%%%%%%%%%%%%%%%%%%%%%%%%%%%%%%%%%%%%%%%%%%
%%%%%%%%%%%%%%%%%%%%%%%%%%%%%%%%%%%%%%%%%%%%%%%%%%%%%%%%%%%%%%%%%%%%%%%%%%%%%%%%%

\bmhead{First-principle modeling of 1L-WSe$_{2}$}
We rely on the tight-binding parametrization of the  spin-dependent DFT band structure provided in Ref.~\cite{liu2013}.
The conduction bands are rigidly shifted upwards by 0.5~eV, to align with the observed quasiparticle bandgap~\cite{madeo2020}.
The eigenvectors ${\bf U}_{\blk }$ of the Bloch Hamiltonian  $H_{{\bf k}}$ are employed to construct the Coulomb integrals entering in the Hartree-SEX and $GW$ self-energies.
These integrals are given by~\cite{wu2015}
$V_{imjn}^{\blq\blk \blk'}=v_{q} (\blU_{ \blk i}^{\dag} \cdot \blU_{ \blk -\blq n})   (\blU_{\blk ' m}^{\dag} \cdot \blU_{ \blk ' +\blq j}) $ 
and describe the scattering amplitude of two electrons initially in bands $j$ and $n$ with momenta $\blk'+\blq$ and $\blk-\blq$ to end up in bands $m$ and $i$ with momenta $\blk'$ and $\blk$ respectively. Here $v_{q}$ represents the Rytova-Keldysh  potential~\cite{keldysh, cudazzo2011} in momentum space, \textit{i.e.},
\be
v_{q}=\frac{2\pi}{ q(1+r_{0}q)} \,
\ee 
where $q=|\blq|$ and $r_{0}=45~\mathring{\rm A}$~\cite{stier2018}. 
The diverging behavior of $v_{q}$ at the $\Gamma$ point is regularized as in Refs.~\cite{huser2013, ridolfi2018}, \textbf{i.e.}, $v(0)\to \frac{1}{\Omega} \int_{\Omega} d\blq \, v(q)$, where $\Omega$ is a 2D domain around $\blq=0$ of linear dimension given by the discretization step of the first Brillouin zone.
From the Bloch Hamiltonian we also calculate the dipole matrix elements for transitions from band $i$ to band $j$ as \cite{perfetto_real_time_2023}
\be
\blD_{ \blk ij}=\frac{1}{i}\frac{1}{\e_{\blk i}-\e_{\blk j}} \blU_{ 
\blk i}^{\dagger}\cdot \partial_{\blk}H_{\blk} \cdot  \blU_{ \blk j}. 
\ee
For the real-time simulations, we restrict our active space to the two highest valence and two lowest conduction bands.
Concerning the phonon input, we parametrize the phonon frequencies as $\w_{ \blk \nu} = \w_{ Q_{\blk} \nu}$, where $Q_{\blk}$ is the high-symmetry point closest to $\blk$.
For the acoustic branches with $\blk$ around  the $\G$ point we instead approximate $\w_{ (\blk \approx 0) \nu}=v_{\nu}|\blk|$. 
The frequencies $\w_{ Q_{\blk} \nu}$ and the sound velocities $v_{\nu}$  are taken from the first-principles calculation reported in Ref.~\cite{li2013}.
Similarly, we approximate the electron-phonon couplings entering the Ehrenfest and FM self-energies as $g_{\nu ij} (\blk,\blq)=\d_{ij} \sqrt{\dfrac{\hbar}{2M\w_{\blq \nu}}} (D^{\nu}_{Q_{\blk} Q_{\blk -\blq} i}+C^{\nu}_{Q_{\blk} Q_{\blk -\blq} i} |\blq|) $, where the intraband zero-th and first order deformation potentials $D^{\nu}_{Q_{\blk} Q_{\blk'}i}$ and $C^{\nu}_{Q_{\blk} Q_{\blk'}i}$ evaluated at the high symmetry points are tabulated in  Ref.~\cite{li2013}.
We consider only the four most coupled %modes LA, TA, LO, TO, namely the 
longitudinal/transverse acoustic/optical phonons.
 
Accurate description of excitons in two-dimensional transition metal dichalcogenides necessitates a dense $\bf{k}$-point discretization of the first Brillouin zone.
However, the numerical solution of Eqs.~(\ref{eom}) in grids having thousands of points is prohibitive.
Given the experimentally observed ultrafast timescales (the EMT occurs within ca.~$70$~fs) we confine the dynamics inside a small plaquette $\mathcal{P}$ around the K-point.
The area  of the plaquette ($A=0.4~\AA \times 0.4~\AA$) and the number of $\blk$-points ($N_{\blk}=169$) are chosen to fulfill two criteria: (i) the solution of the Bethe-Salpeter equation with $\blk\in\callP$ yields the same low-energy spectrum as in the full Brillouin zone, and (ii) the plaquette $\mathcal{P}$ includes all states visited by the carriers upon pumping with the experimental central photon-energy $\w=2.53~$eV.

%%%%%%%%%%%%%%%%%%%%%%%%%%%%%%%%%%%%%%%%%%%%%%%%%%%%%%%%%%%%%%%%%%%%%%%%%%%%%%%%%
%%%%%%%%%%%%%%%%%%%%%%%%%%%%%%%%%%%%%%%%%%%%%%%%%%%%%%%%%%%%%%%%%%%%%%%%%%%%%%%%%
%%%%%%%%%%%%%%%%%%%%%%%%%%%%%%%%%%%%%%%%%%%%%%%%%%%%%%%%%%%%%%%%%%%%%%%%%%%%%%%%%
%%%%%%%%%%%%%%%%%%%%%%%%%%%%%%%%%%%%%%%%%%%%%%%%%%%%%%%%%%%%%%%%%%%%%%%%%%%%%%%%%
%%%%%%%%%%%%%%%%%%%%%%%%%%%%%%%%%%%%%%%%%%%%%%%%%%%%%%%%%%%%%%%%%%%%%%%%%%%%%%%%%

\backmatter

\section*{Supplementary information}

Supplementary Notes~1–10 and Figures~1–10.

\section*{Data availability}

The data supporting the findings of this study are available upon request.

\section*{Acknowledgments}

S.D.C. and G.C. acknowledge financial support by the European Union's NextGenerationEU Programme with the I-PHOQS Infrastructure [IR0000016, ID D2B8D520, CUP B53C22001750006] “Integrated infrastructure initiative in Photonic and Quantum Sciences.” S.D.C. acknowledges support from the European Union's NextGenerationEU – Investment 1.1, M4C2 - Project n. 2022LA3TJ8 – CUP D53D23002280006.
E.P and G.S acknowledge funding from Ministero Universita` e Ricerca PRIN under grant agreement No. 2022WZ8LME, from INFN through project TIME2QUEST, from European Research Council MSCA-ITN TIMES under grant agreement 101118915, and from Tor Vergata University through project TESLA.

\section*{Author contribution}

O.D. and S.D.C. conceived the experiment.
O.D., A.G., and C.T. performed the transient reflectivity measurements. 
O.D. and A.G. analyzed the data. 
A.R.C., J.A.K., E.M.A., and O.B. fabricated the WSe${}_2$ sample and characterized its equilibrium optical response.
K.W. and T.T. grew the hBN crystals.
G.S. and E.P. performed \textit{ab-initio} simulations and theoretical analyses of the experimental results. 
O.D., S.D.C., G.S., and E.P. wrote the manuscript with input from all the authors. 

\section*{Competing interests}

The authors declare no competing interests.

%%%%%%%%%%%%%%%%%%%%%%%%%%%%%%%%%%%%%%%%%%%%%%%%%%%%%%%%%%%%%%%%%%%%%%%%%%%%%%%%%
%%%%%%%%%%%%%%%%%%%%%%%%%%%%%%%%%%%%%%%%%%%%%%%%%%%%%%%%%%%%%%%%%%%%%%%%%%%%%%%%%
%%%%%%%%%%%%%%%%%%%%%%%%%%%%%%%%%%%%%%%%%%%%%%%%%%%%%%%%%%%%%%%%%%%%%%%%%%%%%%%%%
%%%%%%%%%%%%%%%%%%%%%%%%%%%%%%%%%%%%%%%%%%%%%%%%%%%%%%%%%%%%%%%%%%%%%%%%%%%%%%%%%
%%%%%%%%%%%%%%%%%%%%%%%%%%%%%%%%%%%%%%%%%%%%%%%%%%%%%%%%%%%%%%%%%%%%%%%%%%%%%%%%%

\bibliography{sn-bibliography_v2}

\end{document}